\documentclass[amsmath, amssymb, preprintnumbers, showpacs, showkeys,aps,prl,superscriptaddress,twocolumn]{revtex4-2}
\usepackage{booktabs}
\usepackage{graphicx}
\usepackage{dcolumn}
\usepackage{bm}
\usepackage{ bbold }
\usepackage[colorlinks=true,linkcolor=blue,urlcolor=blue,citecolor=blue,pdfusetitle]{hyperref}

\begin{document}

\title{Nonergodic extended phase for waves in three dimensions}

\author{Marcus Prado}
\email{marcus.prado@espci.fr}
\affiliation{Instituto de F\'{\i}sica, Universidade Federal do Rio de Janeiro, Rio de Janeiro-RJ, 21941-972, Brazil}
\affiliation{ESPCI Paris, PSL University, CNRS, Institut Langevin, 75005 Paris, France}
\author{Romain Bachelard}
\affiliation{Universidade Federal de S\~ao Carlos, Rodovia Washington Lu\'{\i}s, km 235-SP-310, 13565-905 S\~ao Carlos, S\~ao Paulo, Brazil}
\author{Robin Kaiser}
\affiliation{Universit\'e C\^ote d’Azur, CNRS, Institut de Physique de Nice, Nice F-06200, France}
\author{Felipe A. Pinheiro}
\affiliation{Instituto de F\'{\i}sica, Universidade Federal do Rio de Janeiro, Rio de Janeiro-RJ, 21941-972, Brazil}

\begin{abstract}
Wave transport in complex media is determined by the nature of quasimodes at the microscopic level. In three dimensional disordered media, waves generally undergo a phase transition from diffusion to Anderson localization, characterized by exponentially localized modes. A remarkable exception are electromagnetic waves, whose vector-like nature prevents Anderson localization to occur. Here we demonstrate that both scalar and vector (electromagnetic) waves exhibit a non-ergodic extended phase characterized by fractal quasimodes, for a broad range of disorder strengths. While electromagnetic waves remain in the non-ergodic extended phase at high disorder strength, scalar waves eventually enter a localized regime. These results pave the way for the engineering of anomalous wave transport phenomena in disordered media without spatial correlations. 
\end{abstract}

\maketitle

\textit{Introduction---} Wave transport in complex media is an ubiquitous phenomenon in nature that underpins a broad range of regimes, both for classical (e.g. electromagnetic, acoustic, seismic) and quantum (matter) waves~\cite{wiersma2013disordered,dalnegro2022waves}. Similar transport regimes occur in different systems, ranging from photonics to condensed matter, atomic physics and acoustics, which puts in evidence the interdisciplinary character of this field. These transport regimes exist in natural media,
such as ballistic and diffusive wave propagation, observed in the atmosphere or in the earth's crust~\cite{wiersma2013disordered}, for example, or they can be engineered in artificial media, such as 
anomalous diffusion in hot atomic clouds~\cite{mercadier2009levy} and in systems with tailored spatial correlations~\cite{bertolotti2010engineering,froufe2017band,dal2017fractional,sellers2017local,aubry2020experimental,yu2021engineered,vynck2023light}, special wavefront-shaped illumination~\cite{rotter2017light,yilmaz2019transverse}, aperiodic photonic metamaterials~\cite{sgrignuoli2020subdiffusive}, turbid media~\cite{pini2024nonselfsimilar}, and fractal structures~\cite{buonsante2011transport}. 

In uncorrelated random media the extreme transport regime in which wave propagation comes to a halt is achieved for strong enough disorder. This phenomenon, known as Anderson localization, has been predicted for electronic propagation in condensed matter as the absence of diffusion due to interference effects~\cite{anderson1958absence}. P.W. Anderson himself realized that, as an undulatory phenomenon, this strong localization can occur for classical waves as well~\cite{anderson1985question}, which has stimulated an intense research activity. Localization of electromagnetic (EM) waves in low dimensions has been reported~\cite{chabanov2000statistical,lahini2008anderson,schwartz2007transport,segev2013anderson,riboli2011anderson}, but in three dimensions (3D), 
its experimental observation has remained elusive despite years of intense research~\cite{wiersma1997localization,scheffold1999localization,storzer2006observation,sperling2013direct,scheffold2013inelastic,sperling2016can,skipetrov2016red}. Indeed, longitudinal EM fields inherent to the vectorial nature of EM waves, and absent from the other (scalar) waves, have been shown to prevent the localization transition in 3D ensembles of randomly distributed point-like scatterers~\cite{skipetrov2014absence,bellando2014cooperative,skipetrov2016finite,skipetrov2018ioffe,cottier2019microscopic,pinheiro2004probing,van2021longitudinal}.
More recently, the suppression of these fields for light scattering in metallic systems allowed for the numerical demonstration of the Anderson localization transition~\cite{yamilov2023anderson,Yamilov2025}.
Alternatively, the modification of level structure of cold atoms due to external fields has been proposed as a viable route to recover Anderson localization for EM waves~\cite{skipetrov2015magneticfielddriven,skipetrov2018localization,celardo2024localization}.

All different transport regimes mentioned above are supported, at the microscopic level, by distinct distributions of spatial eigenstates. For instance, wave diffusion is characterized by delocalized, extended modes, whereas Anderson localization occurs due to exponentially localized eigenstates. In 3D, at the critical point of the Anderson transition the eigenstate amplitudes exhibit strong spatial fluctuations with a multifractal character~\cite{schreiber1991multifractal,rodriguez2010critical,rodriguez2011multifractal}, which leads to anomalous diffusion for scalar waves~\cite{ohtsuki1997anomalous}. In the case of vectorial waves, anomalous light diffusion has been observed in strongly scattering 3D disordered media
~\cite{cobus2022crossover,rezvani2016phase}, with subdiffusion akin to the one predicted by localization theory close to criticality, despite the absence of Anderson localization of light~\cite{cobus2022crossover}. Yet, the underlying EM modal structure which supports this anomalous diffusion remains unknown, which constitutes the first motivation of this work.

Furthermore, recent developments in the realm of many--body localization and random matrix theory have predicted the existence of an entire phase of fractal states~\cite{mace2019multifractal,tarzia2020many,detomasi2021rare,kravtsov2015random,nosov2019correlation}. 
Although these states are extended, they occupy a vanishing portion of the available space in the thermodynamic limit, and are hence called Non-Ergodic Extended (NEE).
In particular, the Rosenzweig-Porter random-matrix ensemble presents another disorder-induced phase transition in addition to the Anderson one: the so-called ergodic transition between ergodic and NEE phases~\cite{kravtsov2015random}.
Fractal phases have also been theoretically predicted in disordered many-body systems~\cite{faoro2019non}, Floquet models~\cite{roy2018multifractality}, the Sachdev-Ye-Kitaev model~\cite{micklitz2019nonergodic}, and graphs~\cite{detomasi2020subdiffusion}. This NEE phase has been shown to be associated with anomalously slow wave dynamics~\cite{khaymovich2021dynamical},
which may find its origin in fractal eigenstates~\cite{ketzmerick1997determines}. Hence, the deviations from conventional diffusion observed in the light propagation 3D in strongly scattering media~\cite{cobus2022crossover,rezvani2016phase} are potential signatures of a NEE phase. 
However, the existence of a NEE phase in a 3D physical model, as well as its impact on transport, remain hardly explored, and this is the second motivation for this work. 

\begin{figure}[t!]
	\centering
    \includegraphics[width=\linewidth]{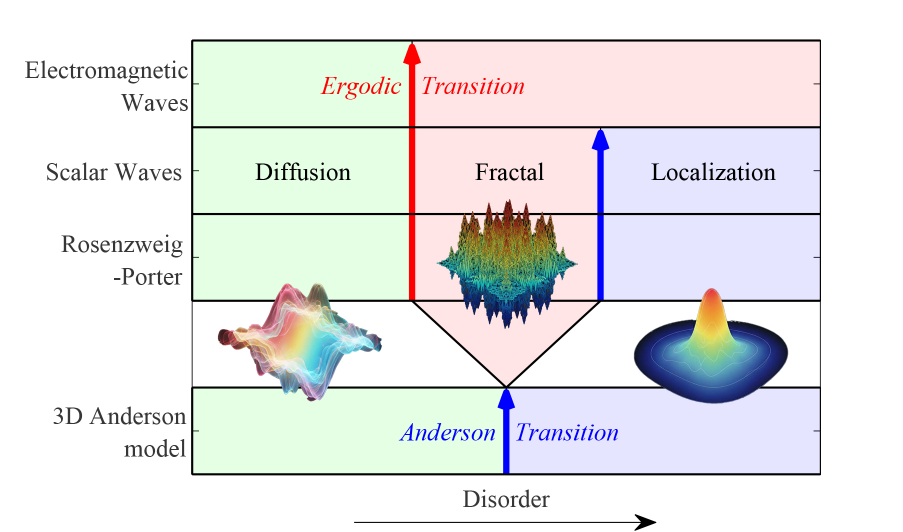}
	\caption{Phase diagram as a function of the disorder strength for the 3D Anderson model, the Rosenzweig-Porter model, and the point-dipole model for scalar and electromagnetic waves with a characteristic, schematic representation of a typical eigenstate of each phase. Vertical arrows indicate the occurence of the Ergodic and Anderson transitions in each model.  }
    \label{fig1}
\end{figure}

In this Letter, we report on the existence of a NEE phase for both scalar and vectorial waves propagating in 3D ensembles of randomly distributed point-like scatterers,
a model that has been experimentally implemented in cold atoms~\cite{labeyrie1999coherent,kupriyanov2003coherent,kwong2014cooperative,araujo2016superradiance,roof2016observation,guerin2016subradiance,bromley2016collective,solano2017superradiance,rui2020subradiant,masson2020manybody,ferioli2021storage,ferioli2024nongaussian}.
Indeed, the spectral analysis of the quasimodes exhibits deviations from the extended, delocalized phase for a finite range of the disorder strength for both scalar and vectorial waves, despite the absence of Anderson localization in the latter case. These deviations, which precede the Ioffe-Regel criterion for localization, are supported at the microscopic level by the presence of fractal eigenstates that emerge after the onset of the ergodic transition. These findings, summarized in Fig.~\ref{fig1}, not only evidence an entire phase of fractal states in a 3D physical model for wave scattering, but it also paves the way for the engineering of novel transport phenomena supported by the fractal phase, for both scalar and electromagnetic waves.

\textit{Microscopic approach---} The existence of the NEE phase for 3D waves is investigated using a microscopic model of randomly distributed point-like scatterers, which has been successfully applied to demonstrate the absence of Anderson localization of EM waves~\cite{skipetrov2014absence} and to characterize the Anderson transition for scalar waves~\cite{skipetrov2016finite,skipetrov2018ioffe}. 
More specifically, we consider dipoles with a single large quality factor resonance (\textit{e.g.}, as in two-level cold atoms) with frequency $\omega_0$ and linewidth $\Gamma_0$. Thus, the collective excitations are described by the effective non-Hermitian Hamiltonian ${\mathcal{H} = \left ( \omega_0 - i\frac{\Gamma_0}{2} \right)\mathbb{1}_{\beta N} - \frac{\Gamma_0}{2}\mathbb{G}}$, where $\beta = 3$ ($\beta = 1$) for EM (scalar) waves and $\mathbb{G}$ is the $\beta N \times \beta N$ Green's matrix that governs the coupling between all $N$ scatterers~\cite{monsarrat2022}. For EM waves, each $3  \times 3$ block is given by the dyadic Green’s function connecting the dipoles located at $\mathbf{r}_i$ and $\mathbf{r}_j$,
\begin{equation}
    \mathbb{G}_{ij} = (1-\delta_{ij})\frac{3e^{ik r_{ij}}}{2kr_{ij}} \left [ P(i k r_{ij}) \mathbb{1}_3 + Q(i k r_{ij}) \frac{\mathbf{r}_{ij} \otimes \mathbf{r}_{ij}}{r_{ij}^2} \right],
\end{equation}
where $k =\omega_0 /c$, $c$ is the speed of light in vacuum,  $\mathbf{r}_{ij} = \mathbf{r}_i - \mathbf{r}_j$, $r_{ij} = |\mathbf{r}_{ij}|$, ${P(z) = 1 - 1/z + 1/z^2}$ and ${Q(z) = -1 +3/z - 3/z^2}$. For scalar waves, the coupling is given by the Green’s function of the Helmholtz equation,
\begin{equation}
    \mathbb{G}_{ij} = (1 - \delta_{ij})\frac{e^{ik r_{ij}}}{kr_{ij}}.
\end{equation}
The complex spectrum of the Green's matrix, denoted as $\{\Lambda_n; n = 1,2,\dots,\beta N\}$, provides information on the frequency positions $\omega_n = \omega_0 - \frac{\Gamma_0}{2}\mathrm{Re}\Lambda_n$ and decay rates $\Gamma_n = \Gamma_0(\mathrm{Im}\Lambda_n+1)$ of the quasimodes~\cite{pinheiro2004probing,skipetrov2014absence}.

We probe the emergence of disorder-induced phase transitions by considering samples with increasing density values $\rho\lambda^3$, where $\lambda = 2\pi/k$. Conversely, the disorder strength decreases with increasing Ioffe-Regel parameter $k\ell$, where $\ell$ is the scattering mean free path. Here, we estimate $\ell$ by its on-resonance value within the independent scattering approximation, namely $\ell = \pi/\rho\lambda^2$ for scalar and $\ell = 2\pi/3\rho\lambda^2$  for EM waves~\cite{carminati2021principles}.

In the following, all analyzed quantities are first averaged within a frequency interval of width $\Gamma_0/2$ around $\omega = \omega_0 + \Gamma_0/4$. Then, an average over $M$ disorder realizations is taken. For all system sizes $N$, $M$ is chosen such that the total number of eigenvalues in the combined spectra $\beta\times N \times M$  is at least $5 \times 10^6$.

\begin{figure}[t!]
	\centering
	\includegraphics[width=\linewidth]{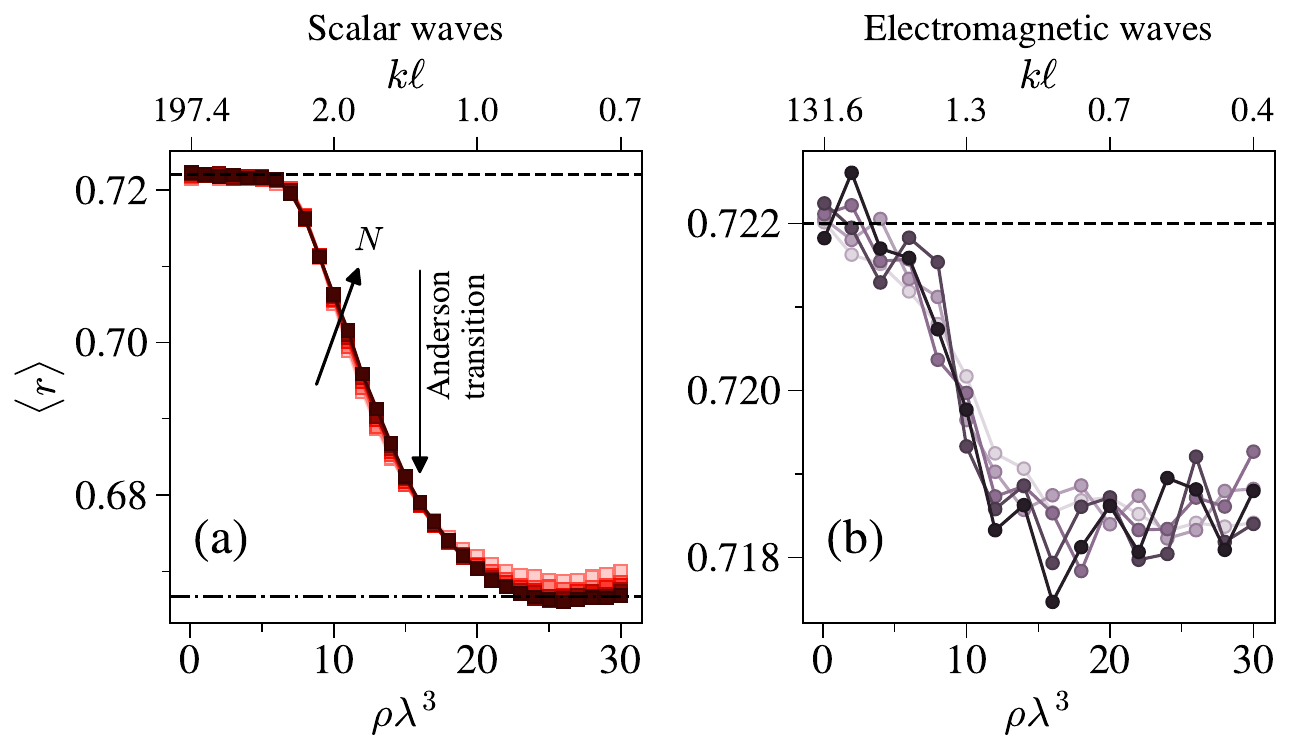}
    \caption{Mean level spacing ratio $\langle r \rangle$ as a function of the scatterer density $\rho\lambda^3$ for (a) scalar and (b) electromagnetic waves. Horizontal dashed (dash-dotted) lines correspond to the predictions for the extended (localized) regime. Darker colors correspond to a larger number of scatterers, whose range is $N \in [2000,12000]$ in (a) and  $N \in [2000,6000]$ in (b). The upper horizontal axes show the Ioffe-Regel parameter $k\ell$ within the independent scattering approximation at resonance. 
    The vertical arrow in (a) shows the location of the Anderson transition.
    }
	\label{fig2}
\end{figure}

\textit{Deviations from spectral signatures of diffusion---} Spectral correlations have been extensively used to characterize the transition between extended and localized phases according to their universal signatures, which depend only on the symmetry class~\cite{evers2008anderson,suntajsSpectralPropertiesThreedimensional2021,romerNumericalMethodsLocalization2024,luo2021universality}. Here, we consider the mean level spacing ratio $\langle r \rangle$~\cite{sa2020complex}. The level spacing ratio is defined as $r_{n} = |\Lambda_{n} - \Lambda_\textrm{NN}| /  |\Lambda_{n} - \Lambda_\textrm{NNN}|$, where $\Lambda_\textrm{NN}$ and $\Lambda_\textrm{NNN}$ are, respectively, the nearest and next-nearest eigenstate to $\Lambda_{n}$ in the complex plane.
Not only is the mean value $\langle r \rangle$ able to determine the critical exponents of the Anderson transition in 3D non-Hermitian disordered systems~\cite{luo2021universality}, but it also does not require unfolding the spectra -- a process prone to numerical artifacts ~\cite{gomez2002misleading,oganesyan2007localization,morales2011improved}. As can be seen in 
Fig.~\ref{fig2}(a), for scalar waves the mean level spacing ratio exhibits a crossing point at a density $\rho_{\mathrm{AT}} \approx 16/\lambda^3$, consistent with the threshold for Anderson localization~\cite{skipetrov2014absence,skipetrov2016finite} and with the Ioffe-Regel criterion $k \ell \approx 1$~\cite{skipetrov2018ioffe}. 
Below (above) the crossover point $\langle r \rangle$ increases (decreases) with the system size $N$, which is consistent with the single-parameter scaling hypothesis~\cite{abrahams1979scaling,suntajsSpectralPropertiesThreedimensional2021,luo2021universality}.

In the diffusive regime, the level statistics of Green's matrices are expected to correspond to those of the $\text{AI}^\dagger$ ensemble of non-Hermitian random matrices, due to their transposition symmetry~\cite{kawabata2019symmetry,garcia-garcia2022symmetry, luo2021universality}. Indeed, at low densities the value $\langle r \rangle_{\mathrm{AI}^\dagger} \approx 0.72$ is obtained for scalar waves~\cite{luo2021universality,garcia-garcia2022symmetry}, as shown in Fig.~\ref{fig2}(a). In contrast, above the density threshold $\langle r \rangle$ reaches the value $2/3$ that corresponds to Poissonian statistics, and it reflects the lack of correlations between the complex frequencies in the localized phase~\cite{sa2020complex,luo2021universality}.

As for EM waves, the diffusive limit is found at low density, see Fig.~\ref{fig2}(b), similar to the low density limit in the scalar case. Surprisingly, the increase in density is also characterized by deviations from the extended phase value $\langle r \rangle_{\mathrm{AI}^\dagger} \approx 0.72$. While Poisson statistics are not expected in the vectorial case due to the absence of the Anderson transition, interestingly the deviations from the value $\langle r \rangle_{\mathrm{AI}^\dagger} \approx 0.72$ occur for the same range of densities for scalar and vectorial waves, that is, from $\rho\lambda^3\approx 8$ to $\approx 20$, for the particle numbers considered. This suggests that even for EM waves, deviations from diffusion occur once the system approaches the Ioffe-Regel criterion $k \ell \approx 1$.

\textit{Fractal dimension of eigenstates---}
Further insights into the origins of these deviations from the extended regime can be better gained from the spatial structure of the eigenstates. The inverse volume occupied by the $n$th state is given by its inverse partipation ratio $\mathrm{IPR}_n = \sum_{i = 1}^N |\psi_n(\mathbf{r}_i)|^4$ \footnote{For all definitions used here, we assume that $\psi$ is normalized as $\sum_{i = 1}^N |\psi(\mathbf{r}_i)|^2 = 1$ and that, for EM waves, $|\psi(\mathbf{r}_i)|$ is treated as a vector norm by taking into account the magnitude of all field polarizations at the position $\mathbf{r}_i$}.
Therefore, the scaling $\langle \mathrm{IPR} \rangle \sim 1/N^{D_F}$ defines the fractal dimension $D_F$ of the states (normalized by the Euclidean dimension)~\cite{evers2008anderson}.
$D_F$ allows a classification in terms of ergodicity, since it determines whether the fraction of occupied volume is unity or vanishes in the thermodynamic limit ($N\to\infty$)~\cite{kravtsov2015random}.

\begin{figure}[t!]
	\centering
	\includegraphics[width=\linewidth]{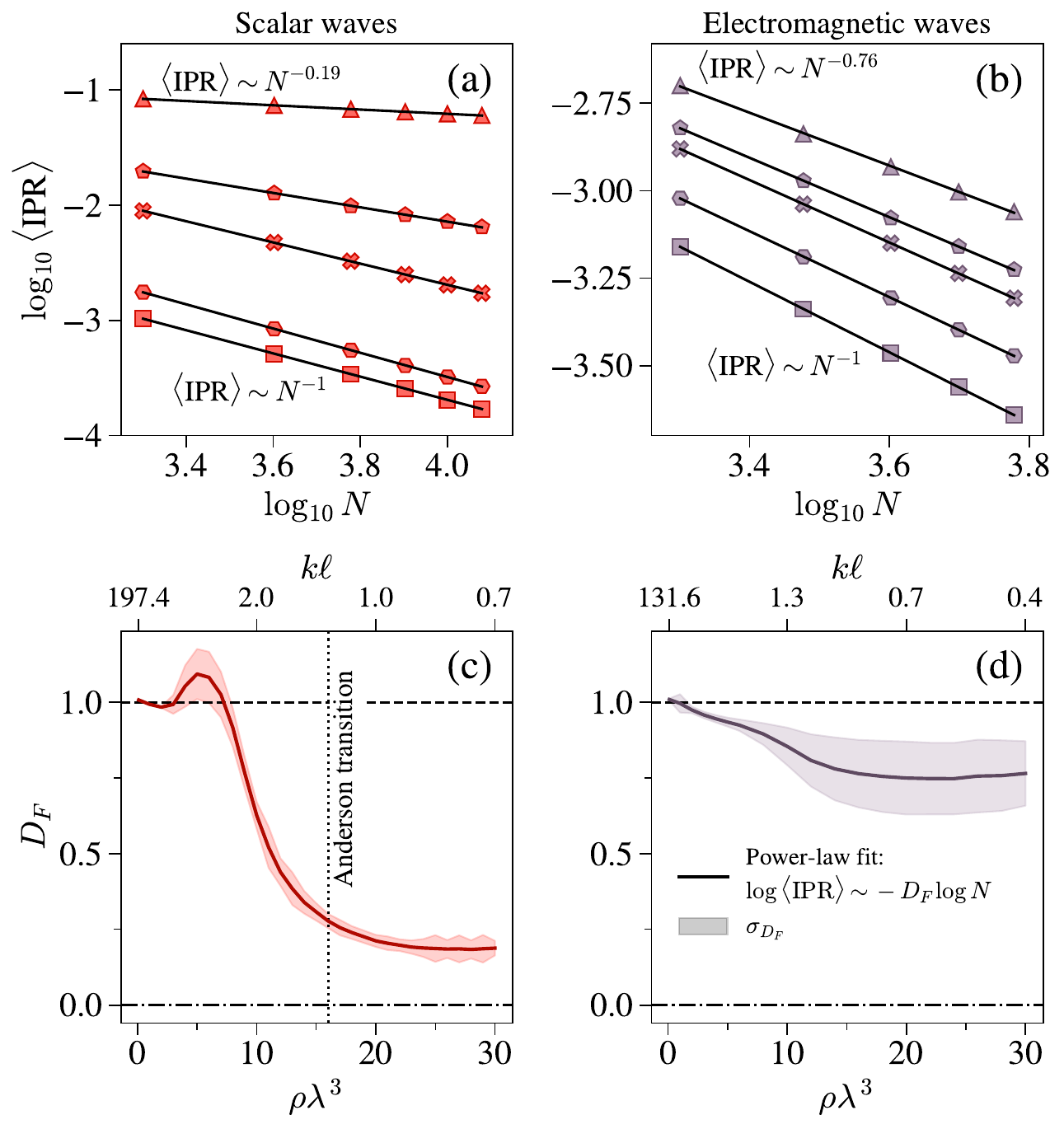}
    \caption{(a),(b) Scaling of the inverse participation ratio (IPR) with system size $N$ for scalar and electromagnetic waves, respectively. 
    Symbols correspond to $\rho\lambda^3 = 0.1,4,8,10,30$, increasing from bottom to top.
    Solid black lines are power-law fits. Panels (c) and (d) present the normalized fractal dimension $D_F$ for the whole range of density. Solid lines show the result of the fitting analysis and the shaded region corresponds to standard deviation. Horizontal dashed (dash-dotted) lines are the predictions for the extended (localized) regime. The vertical line in (a) shows the Anderson transition point extracted from Fig.~\ref{fig2}(a).}
	\label{fig3}
\end{figure}

As shown in Fig.~\ref{fig3}(a-b), a clear power law $\langle \mathrm{IPR} \rangle\sim 1/N^{D_{F}}$ exists for all densities. For both scalar and electromagnetic waves, $D_F\approx 1$ is found at low densities ($\rho \lambda^{3} \lesssim 8 $), which corresponds to the ergodic (\textit{i.e.}, fully extended) phase -- see Fig.~\ref{fig3}(c-d). At large densities ($\rho \lambda^{3} \gtrsim 20$), scalar waves present a fractal dimension $D_{F} \approx 0.2$, not far from the expected $D_{F} = 0$ of the Anderson localized phase. Differently, at high densities vectorial waves exhibit a value $D_q\approx 0.8<1$, which shows that the system is in a NEE (fractal) phase -- see Fig.~\ref{fig3}(d). Hence, while the Thouless number only indicates the absence of the Anderson transition in the vectorial case~\cite{skipetrov2014absence, bellando2014cooperative}, the fractal dimension reveals a change in the nature of eigenstates that support wave transport.

Interestingly, in both scalar and vectorial models a fractal phase emerges at the same density $\rho\lambda^3\approx 8$, below the localization transition $\rho_{\mathrm{AT}} \approx 16/\lambda^3$ for scalar waves, which corresponds to the Ioffe-Regel criterion $k \ell \sim 1$.
Moreover, both types of waves exhibit a crossover in $D_{F}$ within the same density range, namely $\rho\lambda^3 \in [8, 20]$, which closely corresponds to the same range for which spectral deviations of diffusion occurs in $\langle r \rangle$ (see Fig.~\ref{fig2}).

This fractal phase is intermediate for scalar waves since $D_{F}$ eventually drops to a low value at, and beyond, the localization transition threshold $\rho_{\mathrm{AT}}$. Differently, for EM waves $D_{F}$ saturates at a value clearly larger than 0.2 (consistent with the absence of Anderson localization), yet still significantly below 1, which is in turn consistent with non-ergodic, extended states.
This means that scalar waves  eventually localize with increasing densities, whereas vectorial ones remain in a NEE phase. This common range of density for this crossover in the fractal dimension, below the localization threshold, may be due to the fact that the near-field only plays a minor role until the Ioffe-Regel criterion is reached, and thus both types of waves are found in a fractal phase. 

\begin{figure}
    \centering
    \includegraphics[width=\linewidth]{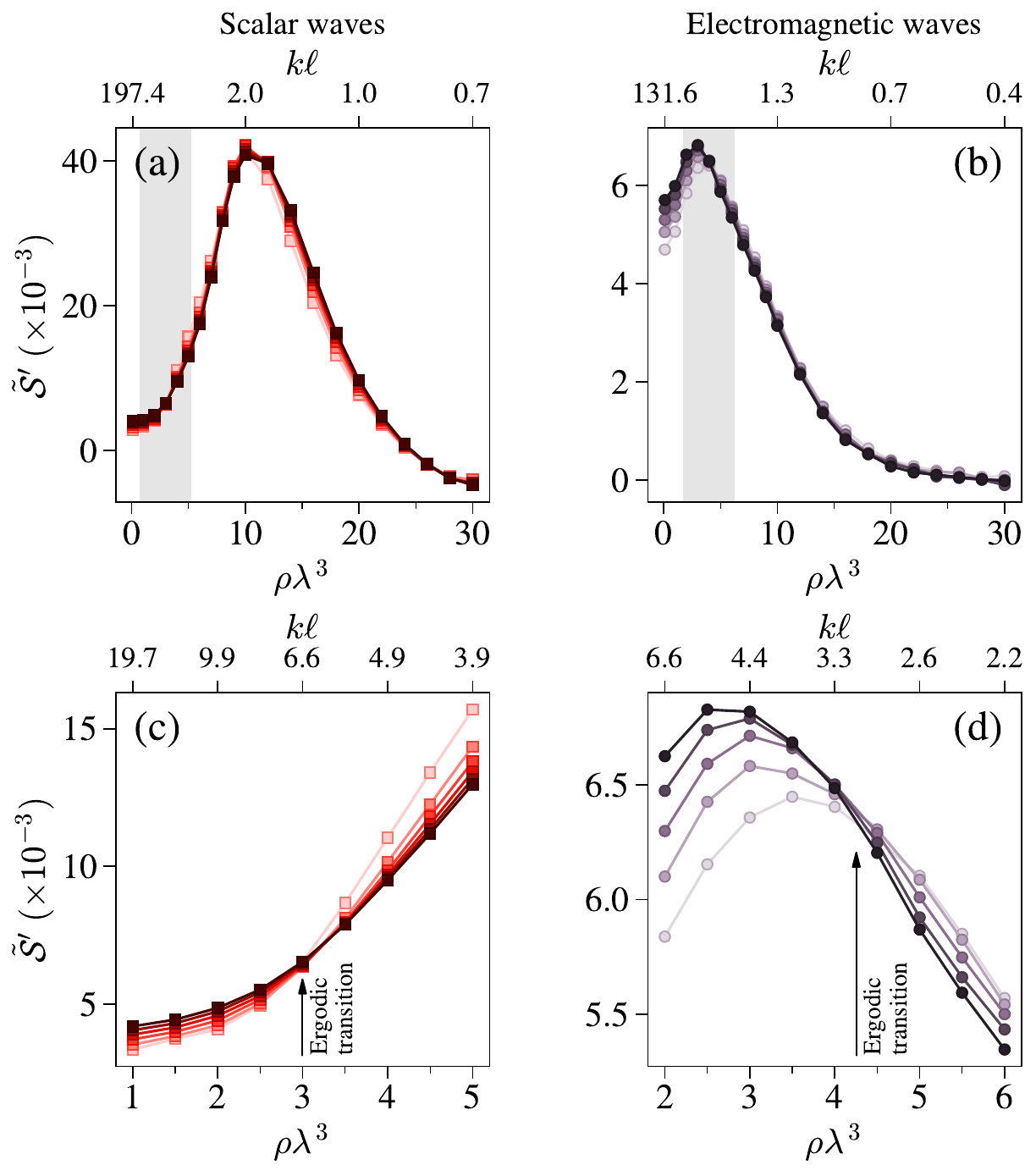}
    \caption{Derivative of the normalized participation entropy $\tilde{S}^{\prime}$ as a function of the scatterer density $\rho\lambda^3$ for (a) scalar and (b) electromagnetic waves. Darker colors correspond to larger system sizes $N$. Panels (c) and (d) show an enlarged view of the curves from (a) and (b) around the crossing points corresponding to the ergodic transition, whose location is indicated by the arrows.}
    \label{fig4}
\end{figure}

\textit{Ergodic transition---}
The existence of the transition to a NEE phase in the thermodynamic limit can be confirmed by a finite-size scaling analysis of the eigenstates.
More specifically, we evaluate the participation entropy $\mathcal{S} =  \langle \sum_{i=1}^N |\psi(\mathbf{r}_i)|^2 \log |\psi(\mathbf{r}_i)|^2 \rangle$ and focus on its normalized derivative $\tilde{\mathcal{S}}^{\prime} = \partial_{\rho\lambda^3} \mathcal{S}/\log N$~\cite{cadez2024rosenzweigporter,Pino2019}.
This approach is known to capture both ergodic and Anderson transitions in the Rosenzweig-Porter model~\cite{cadez2024rosenzweigporter,Pino2019}, in which the occurrence of a NEE phase has been analytically proven~\cite{kravtsov2015random}.
As shown in Fig.~\ref{fig4}, for scalar waves it allows us to identify a new critical point, at a lower density ($ \rho \lambda^{3} \approx 3$), where $\tilde{\mathcal{S}}^{\prime}$ does not depend on the system size, see Fig.~\ref{fig4}(c). It can thus be associated to the ergodic transition between the fully extended and the intermediate NEE phases, as in the Rosenzweig-Porter model~\cite{Pino2019}. As for the vectorial case, a transition point is also found, which occurs at $ \rho \lambda^{3} \approx 4$, see Fig.~\ref{fig4}(d). This again supports the existence of a transition to a NEE phase for electromagnetic waves, despite the absence of the Anderson transition.

\textit{Conclusions---}  We have thus identified a transition to a non-ergodic, extended phase of fractal quasimodes for both scalar and electromagnetic waves propagating in 3D ensembles of randomly distributed point-like scatterers, beyond the simple, traditional scenario of a diffusive-to-localized transition. The direct determination of the fractal dimension reveals a transition to a fractal phase at densities below that of the Anderson  transition for scalar waves. Since fractal states are expected to support anomalous diffusion~\cite{ketzmerick1997determines}, such transport phenomenon could be observed even in a regime where the scaling analysis based on the Thouless number predicts only conventional diffusion.

To the best of our knowledge, this is the first prediction of an NEE phase, where the system crosses over different degrees of (non-)ergodicity for increasing disorder, in a 3D physical model -- as opposed to, for example, the random-matrix based Rosenzweig-Porter model~\cite{cadez2024rosenzweigporter}. Hence, our results may thus be able to not only shed a new light on recent experimental reports of subdiffusive transport of light~\cite{cobus2022crossover,rezvani2016phase}, but also offer further theoretical insights for engineering materials with tailored light transport properties in fully random structures, even without resorting to spatial correlations~\cite{dalnegro2016structural,vynck2023light}, which has been the most common strategy adopted for these purposes so far.

\begin{acknowledgments}
R.K. and F.A.P. thank I. Khaymovich for initial fruitful discussions.
M.P. and F.A.P. acknowledge CNPq, CAPES, and FAPERJ for financial support. F.A.P. acknowledges the financial support of CAPES-COFECUB (Ph 997/23, CAPES 88887.711967/2022-00). R.B. acknowledges the financial support of the São Paulo Research Foundation (FAPESP) (Grants No. 2023/03300-7 and 2022/00209-6), from the Brazilian CNPq (Conselho Nacional de Desenvolvimento Científico e Tecnológico), Grant No. 313632/2023-5 and 403653/2024-0.
R.K. acknowledges funding from ERC Advanced Grant No. 832219 (ANDLICA). 
R.B. and R.K. received support from STIC-AmSud (Ph879-17/CAPES 88887.521971/2020-00), CAPES-COFECUB (Ph 997/23, CAPES 88887.711967/2022-00) and ANR LiLoA (ANR-23-CE30-0035). 
M.P. acknowledges support of the program ``Investissements d’Avenir" launched by the French Government.
\end{acknowledgments}

\end{document}